\documentclass[prd,twocolumn,superscriptaddress,altaffilletter,showpacs,nofootinbib]{revtex4}
\usepackage[dvips]{graphicx}
\usepackage{amsmath}

\newcommand{\be}{\begin{equation}}
\newcommand{\ee}{\end{equation}}
\newcommand{\bea}{\begin{eqnarray}}
\newcommand{\eea}{\end{eqnarray}}
\newcommand{\der}{\partial}

\begin{document}

\title{Introduction to the application of the dynamical systems theory in the study of the dynamics of cosmological models of dark energy}

\author{Ricardo Garc\'{\i}a-Salcedo}\email{rigarcias@ipn.mx}\affiliation{CICATA - Legaria del Instituto Polit\'ecnico Nacional, 11500, M\'exico, D.F., M\'exico.}

\author{Tame Gonzalez}\email{tamegc72@gmail.com}\affiliation{Departamento de Ingenier\'ia Civil, Divisi\'on de Ingenier\'ia, Universidad de Guanajuato, Gto., M\'exico.}

\author{Francisco A. Horta-Rangel}\email{anthort@hotmail.com}\affiliation{Departamento de Ingenier\'ia Civil, Divisi\'on de Ingenier\'ia, Universidad de Guanajuato, Gto., M\'exico.}

\author{Israel Quiros}\email{iquiros6403@gmail.com}\affiliation{Departamento de Ingenier\'ia Civil, Divisi\'on de Ingenier\'ia, Universidad de Guanajuato, Gto., M\'exico.}

\author{Daniel Sanchez-Guzm\'an}\email{dsanchezgzm@gmail.com}\affiliation{CICATA - Legaria del Instituto Polit\'ecnico Nacional, 11500, M\'exico, D.F., M\'exico.}

\date{\today}

\begin{abstract}
The theory of the dynamical systems is a very complex subject which has brought several surprises in the recent past in connection with the theory of chaos and fractals. The application of the tools of the dynamical systems in cosmological settings is less known in spite of the amount of published scientific papers on this subject. In this paper a -- mostly pedagogical -- introduction to the application in cosmology of the basic tools of the dynamical systems theory is presented. It is shown that, in spite of their amazing simplicity, these allow to extract essential information on the asymptotic dynamics of a wide variety of cosmological models. The power of these tools is illustrated within the context of the so called $\Lambda$CDM and scalar field models of dark energy. This paper is suitable for teachers, undergraduate and postgraduate students from physics and mathematics disciplines.
\end{abstract}

\maketitle

%----------------------

\section{Introduction}

Cosmology is about the study of the structure and evolution of the universe as a whole. Einstein's general relativity (GR) stands as its mathematical basis. In general, Einstein's GR equations are a very complex system of coupled non linear differential equations, but it is simplified by the underlying spacetime symmetries. The Friedmann-Robertson-Walker (FRW) spacetime model is depicted by the line element\footnote{For a pedagogical exposition of the geometrical and physical significance of the FRW metric see \cite{sonego}.} $$ds^{2}=-dt^{2}+a(t)^{2}\left[\frac{dr^{2}}{1-k\,r^{2}}+r^{2}(d\theta^{2}+\sin^2\theta d\phi^{2})\right],$$ where $t$ is the cosmic time, $r$, $\theta$ and $\phi$, are co-moving spherical coordinates, $a(t)$ is the cosmological scale factor, and the constant $k=-1,0,+1$ parametrizes the curvature of the spatial sections, is based on what is known as the cosmological principle: in the large scales the Universe is homogeneous and isotropic. Homogeneity and isotropy imply that there is no preferred place and no preferred direction in the universe.\footnote{Here we adopt the units where $8\pi G=c=1$.} The observations confirm the validity of the cosmological principle at large scales ($\sim 100$ Mpc $\approx 10^{26}$ cm). The dynamics of the universe is fully given by the explicit form of the scale factor which depends on the symmetry properties and on the matter content of the Universe. 

In order to determine the scale factor it is necessary to solve the cosmological Einstein's field equations, which relate the spacetime geometry with the distribution of matter in the Universe. The maximal symmetry implied by the cosmological principle allows for great simplification of the Einstein's equations. In this paper, for further simplification, we shall consider FRW spacetimes with flat spatial sections ($k=0$). The resulting cosmological equations are written as it follows:\footnote{For a didactic derivation of the cosmological equations not based on general relativity see \cite{sonego, uzan, other}.}

\bea &&3H^2=\sum_x\rho_x+\rho_X,\nonumber\\
&&\dot H=-\frac{1}{2}\left(\sum_x\gamma_x\rho_x+\gamma_X\rho_X\right),\nonumber\\
&&\dot\rho_x+3\gamma_x H\rho_x=0,\;\dot\rho_X+3\gamma_X H\rho_X=0,\label{x-cons-eq}\eea where $H\equiv\dot a/a$ is the Hubble parameter, $\rho_x$ and $\rho_X$ are the energy densities of the gravitating matter and of the $X$-fluid (the dark energy in the present paper) respectively, and the sum is over all of the gravitating matter species living in the FRW spacetime (dark matter, baryons, radiation, etc.). The last two equations above are the conservation equations for the gravitating matter degrees of freedom and for the $X$-fluid respectively.\footnote{It should be pointed out that one equation in (\ref{x-cons-eq}), say the second (Raychaudhuri) equation, is redundant.} Besides, it is adopted that the following equations of state are obeyed: $$p_x=(\gamma_x-1)\rho_x,\;p_X=(\gamma_X-1)\rho_X,$$ where $p_x$ is the barotropic pressure of the $x$-matter fluid, $p_X$ is the parametric pressure of $X$-component,\footnote{By parametric pressure we understand that it might not correspond to standard barotropic pressure with the usual thermodynamic properties but that it is rather a conveniently defined parameter which obeys similar equations than its matter counterpart.} while $\gamma_x$ and $\gamma_X$ are their barotropic parameters.\footnote{In several places in this paper we use the so called equation of state (EOS) parameter $\omega_X=p_X/\rho_X=\gamma_X-1$, interchangeably with the barotropic parameter $\gamma_X$.}

According to our current understanding of the expansion history, the Universe was born out of a set of initial conditions known as ``big-bang.'' Further evolution led to a stage of primeval inflation driven by a scalar field \cite{liddle}. The inflation era was followed by a stage dominated by the radiation, followed, in turn, by an intermediate stage dominated by non-relativistic cold dark matter (CDM).\footnote{Dark matter is a conventional form of matter in the sense that it gravitates as any other known sort of standard matter such as radiation, baryons, etc. However, unlike the latter forms of matter, CDM does not interact with radiation. This is why it is called as ``dark matter.''} During this stage of the cosmic expansion most of the structure we observe was formed. The end of the matter-domination era can be traced back to a recent moment of the cosmic expansion where a very peculiar form of matter which does not interact with baryons, radiation or any other form of ``visible'' matter -- known as ``dark energy'' (DE) -- started dominating \cite{copeland-rev}. This alien form of matter antigravitates and causes the present Universe to expand at an accelerating pace instead of decelerating, as it would for a matter -- either CDM or baryons -- or radiation dominated Universe. Actually, if one substitutes $$\frac{\ddot a}{a}=\dot H+H^2,$$ back into the second equation in (\ref{x-cons-eq}), and taking into account the Friedmann equation -- first equation in (\ref{x-cons-eq}) -- then, one is left with the following equation:

\bea \frac{\ddot a}{a}=-\frac{1}{2}\sum_x\left(\gamma_x-\frac{2}{3}\right)\rho_x-\frac{1}{2}\left(\gamma_X-\frac{2}{3}\right)\rho_X.\label{accel-eq}\eea 

It is known that for standard (gravitating) forms of matter the barotropic pressure is a non-negative quantity: $p_x=(\gamma_x-1)\rho_x\geq 0$ $\Rightarrow\;\gamma_x\geq 1$. In particular, for CDM and baryons the (constant) barotropic index $\gamma_m=1$, while for radiation $\gamma_r=4/3$. Hence, since obviously $\gamma_x>2/3$ -- if forget for a while about the second term in the RHS of (\ref{accel-eq}) -- it is seen that a cosmic background composed of standard matter will expand at a decelerated pace ($\ddot a<0$). In contrast, for unconventional matter with $\gamma_X<2/3$ (the parametric pressure $p_X$ is obviously negative), the second term in the RHS of Eq. (\ref{accel-eq}) contributes towards accelerating the pace of the cosmological expansion. In particular, if we choose the $X$-component in the form of vacuum energy $\left(\gamma_\text{vac},\rho_\text{vac}\right)\rightarrow\left(\gamma_X,\rho_X\right)$, usually identified with the energy density of the cosmological constant (see section \ref{lcdm}), since $p_\text{vac}=-\rho_{vac}$, then $\gamma_\text{vac}=0<2/3$. Due to the known fact that the vacuum energy is not diluted by the cosmic expansion,\footnote{We recall that, in contrast with the vacuum energy density which does not dilute with the expansion $\rho_\text{vac}=\Lambda=const.$, the energy density of CDM and baryons dilutes like $\propto a^{-3}$, while the density of radiation decays very quickly $\propto a^{-4}$.} eventually this component of the cosmic budget will dominate the cosmological dynamics, leading to late-time acceleration of the expansion $$\frac{\ddot a}{a}\approx \frac{1}{3}\,\rho_\text{vac}>0.$$ 

A related (dimensionless) parameter which measures whether the expansion is accelerated or decelerated, is the so called deceleration parameter which is defined as

\bea q\equiv-1-\frac{\dot H}{H^2}=-\frac{a\ddot a}{\dot a^2}.\label{q-def}\eea The pace of the expansion accelerates if $q<0$. On the contrary, positive $q>0$ means that the expansion is decelerating. This parameter will be useful in the discussion in the next sections.

In spite of the apparent simplicity of the equations (\ref{x-cons-eq}), as with any system of non-linear second-order differential equations, it is a very difficult (and perhaps unsuccessful) task to find exact solutions. Even when an analytic solution can be found it will not be unique but just one in a large set of them \cite{faraoni}. This is in addition to the question about stability of given solutions. In this case the dynamical systems tools come to our rescue. These very simple tools give us the possibility to correlate such important concepts like past and future attractors (also saddle equilibrium points) in the phase space, with generic solutions of the set of equations (\ref{x-cons-eq}) without the need to analytically solve them. 

In correspondence with the above mentioned stages of the cosmic expansion, one expects that the state space (also, phase space) of any feasible cosmological model should be characterized by -- at least one of -- the following equilibrium configurations in the space of cosmological states: (i) a scalar field dominated (inflationary) past attractor, (ii) saddle critical points associated with radiation-domination, radiation-matter scaling, and matter-domination, and either (iii) a matter/DE scaling or (iv) a de Sitter\footnote{A de Sitter cosmological phase is a stage of the cosmological expansion whose dynamics is described by the line element $ds^2=-dt^2+e^{H_0t}d{\bf x}^2$, where $H_0$ is the Hubble constant. Since the deceleration parameter $q\equiv-1-\dot H/H^2=-1$ is a negative quantity, the de Sitter expansion is inflationary.} future attractors. Any heteroclinic orbit joining these critical points then represents a feasible image of the featured cosmic evolution in the phase space.\footnote{For a very concise introductory knowledge of the theory of the dynamical systems and the related concepts like critical points (attractors and saddle points), heteroclinic orbits, etc, see the appendix \ref{appendix}.} 

While the theory of the dynamical systems has brought several surprises in the recent past in connection with the theory of chaos \cite{chaos, hirsh} and fractals \cite{fractals}, the application of the tools of the dynamical systems in cosmological settings is less known in spite of the amount of published scientific papers on this subject (just to cite few of them see \cite{uzan, copeland-rev, ellis, wands, coley, amendola, nunez, quiros-prd-2009, luis, quiros, lazkoz, boehmer} and the related references therein). 

This paper aims at an introduction to the application of the simplest tools of the theory of the dynamical systems -- those which can be explained with only previous knowledge of the fundamentals of linear algebra and of ordinary differential equations -- within the context of the so called ``concordance'' or $\Lambda$CDM (section \ref{lcdm}) and scalar field models (section \ref{scalar-f}) of dark energy \cite{copeland-rev, wands, coley, amendola, nunez, quiros-prd-2009, luis, lazkoz, boehmer, mimoso, barreiro, linder}. The dynamical systems tools play a central role in the understanding of the asymptotic structure of these models. I. e., these tools are helpful in looking for an answer to questions like: where does our Universe come from? or, what would be its fate? An specific example: the cosh potential, is explored in the section \ref{cosh-sect}, in order to show the power of these tools in the search for generic dynamical behavior.

We want to stress that there are very good introductory and review papers on the application of the dynamical systems in cosmology -- see, for instance, \cite{copeland-rev, ellis, wands, coley, luis} -- so, what is the point of writing another introductory paper on the issue? In this regard we want to point out that, in the present paper, we aim not only at providing just another introductory exposition of the essentials of the subject but, also, at filling several gaps regarding specific topics usually not covered in similar publications existing in the bibliography. Here we pay special attention, in particular, to the necessary (yet usually forgotten) definition of the physically meaningful phase space, i. e., that region of the phase space which contains physically meaningful cosmological solutions. We also include the derivation of relevant formulas in order to facilitate the way for beginners to concrete computations. We shall explain, in particular, a method developed in \cite{z} to consider arbitrary potentials within the dynamical systems study of scalar field cosmological models. This subject is not usually covered in the existing introductory bibliography. At the end of the paper, for completeness, we include an appendix section \ref{appendix}, where an elementary introduction to the theory of the dynamical systems is given and useful comments on the interplay between the cosmological field equations and the equivalent phase space are also provided. Only knowledge of the fundamentals of linear algebra and of ordinary differential equations is required to understand the material in this section. The exposition in the appendix is enough to understand how the computations in sections \ref{lcdm} and \ref{scalar-f} are done. We recommend those readers which are not familiar with the dynamical systems theory to start with this section. 

Our goal is to keep the discussion as general as possible while presenting the exposition in as much as possible pedagogical way. The paper contains many footnotes with comments and definitions which complement the main text. Since the subject is exposed with some degree of technical details and a basic knowledge of cosmology is assumed, this paper is suitable for teachers, undergraduate and postgraduate students from physics and mathematics disciplines.

%------------------------------------------------

\section{Asymptotic cosmological dynamics: a model-independent analysis}\label{model-indep}

In this section the asymptotic structure in the phase space of a general DE model will be discussed in detail. For simplicity and compactness of the exposition here we shall assume only a two-component cosmological fluid composed of cold dark matter -- labeled here by the index ``$m$,'' i. e., $(\rho_m,p_m)\rightarrow(\rho_x,p_x)$ -- and of dark energy (the $X$-component). 

In order to derive an autonomous system of ordinary differential equations (ASODE) out of (\ref{x-cons-eq}) it is useful to introduce the so called dimensionless energy density parameters of matter and of the DE

\bea \Omega_m=\frac{\rho_m}{3H^2},\;\Omega_X=\frac{\rho_X}{3H^2},\label{dens-par}\eea respectively, which are always non-negative quantities.\footnote{In agreement with conventional non-negativity of energy we consider non-negative energy densities exclusively.} In terms of these parameters the Friedmann equation -- first equation in (\ref{x-cons-eq}) -- can be written as the following constraint:

\bea \Omega_m+\Omega_X=1,\label{friedmann-c}\eea which entails that none of the non-negative dimensionless energy density components alone may exceed unity: $0\leq\Omega_m\leq 1$, $0\leq\Omega_X\leq 1$. 

Given the above constraints, if one thinks of the dimensionless density parameters as variables of some state space, only one of them is linearly independent, say $\Omega_X$. Let us write a dynamical equation for $\Omega_X$, for which purpose we rewrite the Raychaudhuri equation -- second equation in (\ref{x-cons-eq}) -- and the conservation equation for the $X$-component, in terms of $\Omega_X$:

\bea &&2\frac{\dot H}{H^2}=-3\gamma_m\Omega_m-3\gamma_X\Omega_X\nonumber\\
&&\;\;\;\;\;\;\;\;=-3\gamma_m+3(\gamma_m-\gamma_X)\Omega_X,\nonumber\\
&&\frac{\dot\rho_X}{3H^2}=-3\gamma_X H\Omega_X.\label{dot-h}\eea Then, if we substitute (\ref{dot-h}) into $$\dot\Omega_X=\frac{\dot\rho_X}{3H^2}-2\frac{\dot H}{H}\,\Omega_X,$$ and consider the constraint (\ref{friedmann-c}), we obtain

\bea \Omega'_X=3(\gamma_m-\gamma_X)\Omega_X(1-\Omega_X),\label{ode-1}\eea where the tilde denotes derivative with respect to the new variable $\tau\equiv\ln a$ ($\dot\Omega=H\Omega'$, etc.). 

In general, for varying $\gamma_X$, another ordinary differential equation (ODE) for $\gamma_X$ is needed in order to have a closed system of differential equations (the barotropic index of matter $\gamma_m$ is usually set to a constant). The problem is that, unlike the ODE (\ref{ode-1}) which is model-independent, the ODE for $\gamma_X=(\rho_X+p_X)/\rho_X$ requires of certain specifications which depend on the chosen model (see below). 

Anyway, certain important results may be extracted from (\ref{ode-1}) under certain assumptions. For instance, if we assume a constant $\gamma_X$, the mentioned ODE is enough to uncover the asymptotic structure of the model (\ref{x-cons-eq}) in the phase space. In this case the phase space is the 1-dimensional segment $\Omega_X\in [0,1]$. The critical points are $$\Omega_X(1-\Omega_X)=0\;\Rightarrow\;\Omega_X=1,\;\Omega_X=0,$$ where we are assuming that $\gamma_X\neq\gamma_m$. The first fixed point $\Omega_X=1$ is correlated with dark energy domination, while the second one $\Omega_X=0$ ($\Omega_m=1$) is associated with CDM dominated expansion.

If linearize Eq. (\ref{ode-1}) around the critical points: $1-\delta_1\rightarrow 1$, and $0+\delta_0\rightarrow 0$ ($\delta_1$ and $\delta_0$ are small perturbations), one obtains

\bea &&\delta'_1=-3(\gamma_m-\gamma_X)\delta_1\;\Rightarrow\;\delta_1(\tau)=\bar\delta_1\,e^{-3(\gamma_m-\gamma_X)\tau},\nonumber\\
&&\delta'_0=3(\gamma_m-\gamma_X)\delta_0\;\Rightarrow\;\delta_0(\tau)=\bar\delta_0\,e^{3(\gamma_m-\gamma_X)\tau}.\label{sol-ode-1}\eea 

As seen the DE-dominated solution: $$\Omega_X=1\;\Rightarrow\;3H^2=\rho_X\propto a^{-3\gamma_X}\;\Rightarrow\;a(t)\propto t^{2/3\gamma_X},$$ is a stable (isolated) equilibrium point whenever $\gamma_m>\gamma_X$ since, in this case, the perturbation $$\delta_1(\tau)\propto e^{-3(\gamma_m-\gamma_X)\tau},$$ decreases exponentially. Hence, given that $\gamma_m>\gamma_X$, the dark energy dominated solution $\Omega_X=1$, is the future attractor solution which describes the fate of the Universe. Besides, if $\gamma_m>\gamma_X$, the matter-dominated solution $\Omega_m=1$ ($\Omega_X=0$): $a(t)\propto t^{2/3\gamma_m},$ is the past attractor since the initially small perturbation $\delta_0(\tau)\propto e^{3(\gamma_m-\gamma_X)\tau}$ increases exponentially with $\tau$, thus taking the system away from the condition $\Omega_m=1$. Otherwise, if $\gamma_m<\gamma_X$, the $X$-dominated solution is the past attractor while the matter-dominated solution is the future attractor. However, this last situation is not consistent with the known cosmic history of our Universe.

The interesting thing about the above result is that it is model-independent, i. e., no matter which model to adopt for the DE ($X$-component), given that $\gamma_X$ is a constant, the above mentioned critical points -- together with their stability properties -- are obtained. Besides, since a non-constant $\gamma_X$ obeys an autonomous ODE of the general form $\gamma'_X=f(\gamma_X,\Omega_X)$, at any equilibrium point where, necessarily $$\Omega'_X=0,\;\gamma'_X=0\;\Rightarrow\;\gamma_X=\bar\gamma_X=const.$$ provided that $\bar\gamma_X\neq\gamma_m$, either $\Omega_X=1$, or $\Omega_X=0$ ($\Omega_m=1$). Hence, independent of the model adopted to account for the $X$-component, matter-dominated and $X$-fluid dominated solutions are always equilibrium points in the phase space of the two-fluids cosmological model depicted by equations (\ref{x-cons-eq}). This is a generic result.

\section{$\Lambda$CDM model}\label{lcdm}

Let us now to show with the help of a specific cosmological model of dark energy, how it is possible that a 2-dimensional system of ODE may come out of the 3-dimensional system of second-order field equations (\ref{x-cons-eq}) where, we recall, one of these equations is redundant. For this purpose we shall focus in the so called $\Lambda$-cold dark matter ($\Lambda$CDM) model \cite{lcdm-bib} whose phase space is a subspace of the phase plane.\footnote{For a dynamical systems study of the FRW cosmological model with spatial curvature $k\neq 0$ and with a cosmological constant see \cite{uzan}.} Here -- only for the extent of this section -- in addition to the CDM and to the DE (the cosmological constant) we shall consider a radiation fluid with energy density $\rho_r$ and pressure $p_r=\rho_r/3$ ($\gamma_r=4/3$). The conservation equation for radiation reads $$\dot\rho_r+4H\rho_r=0\;\Rightarrow\;\rho_r\propto a^{-4}.$$ Meanwhile, for the cosmological constant term -- as for any vacuum fluid -- we have that $p_\Lambda=-\rho_\Lambda$ ($\gamma_\Lambda=0$), which means that $\dot\rho_\Lambda=0$, so that this term does not evolve during the course of the cosmic expansion. 

In what follows, since we adopted the units system with $8\pi G=c=1$, we write $\rho_\Lambda=\Lambda$. Besides, since we deal here with CDM, which is modeled by a pressureless dust, we set $\gamma_m=1$. After these assumptions, and identifying $\rho_X\equiv\rho_\Lambda=\Lambda$ in equations (\ref{x-cons-eq}), the resulting cosmological equations are

\bea &&3H^2=\rho_r+\rho_m+\Lambda,\;\dot H=-\frac{1}{2}\left(\frac{4}{3}\rho_r+\rho_m\right),\nonumber\\
&&\dot\rho_m+3H\rho_m=0,\;\dot\rho_r+4H\rho_r=0.\label{r-cons-eq'}\eea 

Let us at this point to make a step aside to explain how it is that the cosmological constant can explain the present speed-up of the cosmic expansion. If we conveniently combine the first two equations above, taking into account that straightforward integration of the last equations in (\ref{r-cons-eq'}) yield $\rho_m=6C_m/a^3$ and $\rho_r=3C_r/a^4$ respectively ($C_m$ and $C_r$ are constants), we get: $$\frac{\ddot a}{a}=\dot H+H^2=-\frac{C_r}{a^4}-\frac{C_m}{a^3}+\frac{\Lambda}{3}.$$ 

We see that acceleration of the expansion $\ddot a>0$ may occur in this model thanks to the third term above (the DE, i. e., the cosmological constant). Besides, the accelerated pace of the expansion is a recent phenomenon: during the course of the cosmic evolution the initially dominating radiation component dilutes very quickly $\propto a^{-4}$ until the matter component starts dominating. As the cosmic expansion further proceeds the matter component also dilutes $\propto a^{-3}$ until very recently when the cosmological constant started to dominate to yield to positive $\ddot a>0$.

Following the same procedure as in the former section we write the Friedmann constraint 

\bea \Omega_r+\Omega_m+\Omega_\Lambda=1,\label{friedmann-c'}\eea where $\Omega_r=\rho_r/3H^2$ and $\Omega_\Lambda=\Lambda/3H^2$. In what follows we choose the following variables of the phase space:

\bea x\equiv\Omega_r,\;y\equiv\Omega_\Lambda,\label{rad-vars}\eea where, for sake of simplicity of writing, we adopt $x$ and $y$ in place of the dimensionless energy densities of radiation $\Omega_r$ and of the cosmological constant $\Omega_\Lambda$ respectively. But we warn the reader that the same symbols $x$ and $y$ will be used in the next sections to mean different variables of the phase space. 

\begin{figure}[t!]\begin{center}
\includegraphics[width=5cm]{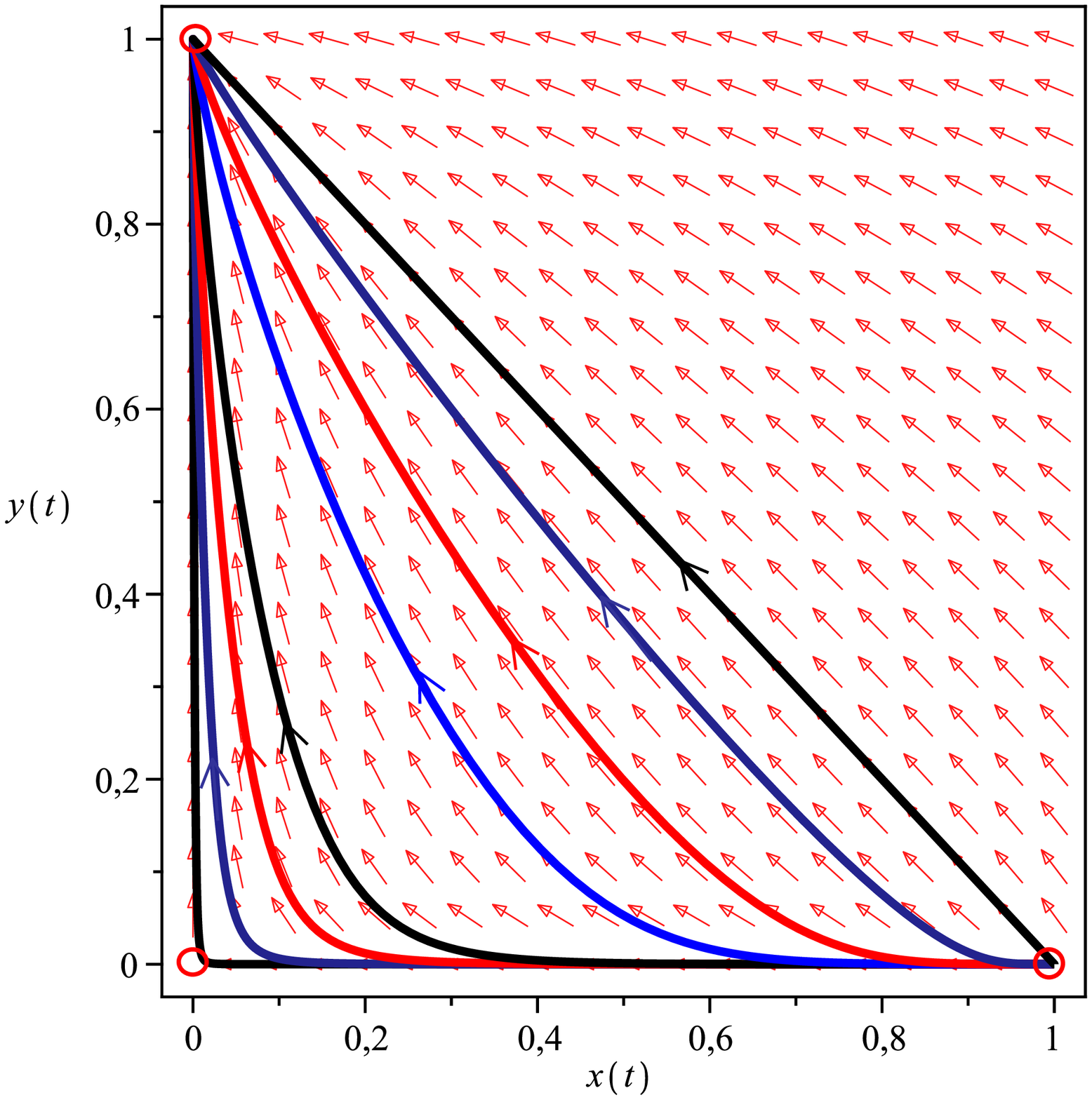}
\includegraphics[width=5cm]{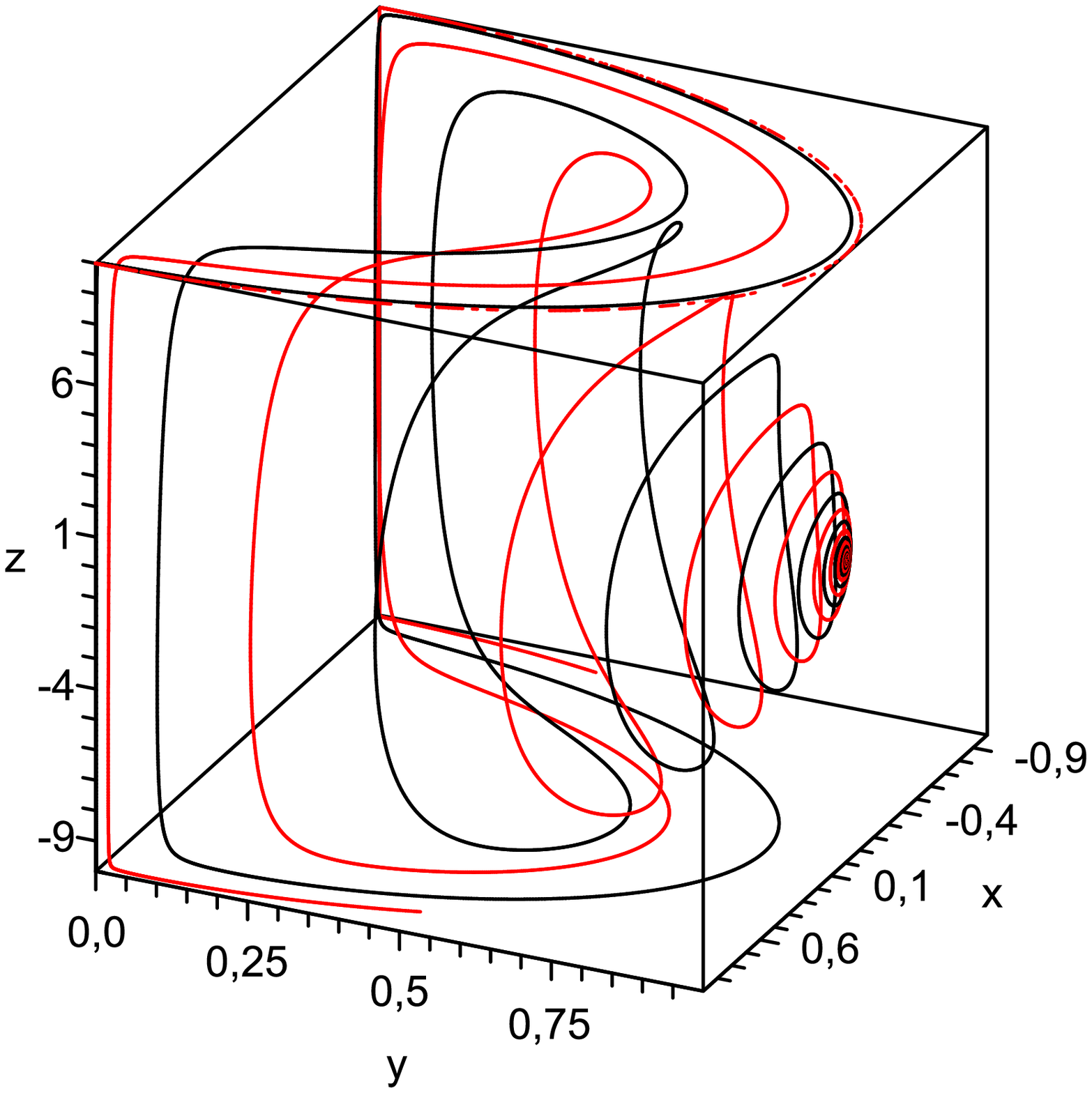}\vspace{0.7cm}
\caption{Phase portrait of the ASODE (\ref{r-ode}) corresponding to the $\Lambda$CDM model (top panel). The orbits of (\ref{r-ode}) emerge from the past attractor $P_r:(1,0)$ -- the radiation domination state -- and end up in the future attractor $P_{dS}:(0,1)$ (de Sitter phase). For a non-empty set of initial conditions the corresponding orbits approach close enough to the matter-dominated saddle point at the origin $P_m:(0,0)$. In the bottom panel the phase portrait of the ASODE (\ref{cosh-ode}) -- showing two orbits generated by two different sets of initial conditions -- is shown. Here the free parameter $\mu$ has been arbitrarily set $\mu=10$.}\label{fig-lcdm}\end{center}
\end{figure}

We have that $\Omega_m=1-x-y$, and, since $0\leq x\leq 1$, $0\leq y\leq 1$, and $0\leq\Omega_m\leq 1$, the physically relevant phase space is defined as the following 2D triangular region (see the top panel of FIG. \ref{fig-lcdm}):

\bea \Psi_\Lambda=\{(x,y):\,x+y\leq 1,\;0\leq x\leq 1,\;0\leq y\leq 1\}.\label{psi-l}\eea 

As before, in order to derive the autonomous ODE-s, we first take the derivative of the variables $x$ and $y$ with respect to the cosmic time $t$: $$\dot x=\frac{\dot\rho_r}{3H^2}-2\frac{\rho_r}{3H^2}\frac{\dot H}{H},\;\dot y=-2\frac{\Lambda}{3H^2}\frac{\dot H}{H},$$ where we have to take in mind the Raychaudhuri equation and the conservation equation for radiation in (\ref{r-cons-eq'}) written in terms of the variables of the phase space $$\dot H=-\frac{H^2}{2}\left(3+x-3y\right),$$ and then we have to go to derivatives with respect to the variable $\tau=\ln a(t)$ ($\xi'\rightarrow H^{-1}\dot\xi$). We obtain:

\bea x'=-x\,(1-x+3y),\;y'=(3+x-3y)\,y.\label{r-ode}\eea

It is remarkable the simplicity of the system of two ordinary differential equations (\ref{r-ode}) as compared with the system of three second-order cosmological equations (\ref{r-cons-eq'}). The critical points of (\ref{r-ode}) in $\Psi_\Lambda$ are easily found by solving the following system of algebraic equations: $x\,(1-x+3y)=0,\;(3+x-3y)\,y=0$.

In the present case, in terms of the variables $x$, $y$, the deceleration parameter $q=-1-\dot H/H^2$ (equation (\ref{q-def})) can be written as: $$q=(1+x-3y)/2,$$ which means that those fixed points which fall in that region of the phase space laying above the line $y=1/3+x/3$, correspond to cosmological solutions where the expansion is accelerating.

%---------------------------------------

%----------------------------------------

\subsection{Critical points}

\subsubsection{Radiation-dominated critical point}

The critical point $P_r:(1,0),\;\Omega_r=1$ $\Rightarrow\;3H^2=\rho_r$ $\Rightarrow\;a(t)\propto\sqrt{t}$, corresponds to the radiation-dominated cosmic phase. This solution depicts decelerated expansion since $q=1$. The linearization matrix for the system of ODE (\ref{r-ode}) -- for details see the appendix -- is $$J=\begin{pmatrix} -1+2x-3y & -3x \\ y & 3+x-6y \end{pmatrix}.$$ The roots of the algebraic equation $$\det|J(P_r)-\lambda U|=\det\begin{pmatrix} 1-\lambda & -3 \\ 0 & 4-\lambda \end{pmatrix}=0,$$ are the eigenvalues $\lambda_1=1$ and $\lambda_2=4$. Since both eigenvalues are positive reals, then $P_r$ is a source point (past attractor). This means that in the $\Lambda$CDM model described by the cosmological equations (\ref{r-cons-eq'}) the radiation-dominated solution $a(t)\propto\sqrt{t}$ is a privileged solution. 

From the inspection of the phase portrait of (\ref{r-ode}) -- top panel of FIG. \ref{fig-lcdm} -- it is apparent that any viable pattern of cosmological evolution should start in a state where the matter content of the Universe is dominated by radiation. Of course this is a drawback of our classical model which can not explain the very early stages of the cosmic evolution where quantum effects of gravity play a role. This includes the early inflation. In a model that would account for such early period of the cosmic dynamics, inflation should be the past attractor in the equivalent phase space.

\subsubsection{Matter-dominated critical point}

The critical point $P_m:(0,0),\;\Omega_m=1$ $\Rightarrow\;3H^2=\rho_m$ $\Rightarrow\;a(t)\propto t^{2/3}$, corresponds to the matter-dominated solution, which is associated with decelerated expansion ($q=1/2$). Following the same procedure above -- see the appendix \ref{appendix} -- one finds the following eigenvalues of the linearization matrix $J(P_m)$ evaluated at the hyperbolic equilibrium point $P_m$: $\lambda_1=1$, $\lambda_2=-3$. Hence this is a saddle critical point. As seen from the inspection of the phase portrait in the top panel of FIG. \ref{fig-lcdm}, only for conveniently chosen initial conditions the given orbits in $\Psi_\Lambda$ approach close enough to $P_m$. Since this is a unstable (metastable) critical point, it can be associated with a transient stage of the cosmic expansion only. This is good since a stage dominated by the dark matter can be only a transient state, lasting for enough time as to account for the observed amount of cosmic structure. A drawback is that we need to fine tune the initial conditions in order for feasible orbits in the phase space to get close enough to $P_m$.

\subsubsection{de Sitter phase}

The critical point $P_{dS}:(0,1),\;\Omega_\Lambda=1$ $\Rightarrow\;3H^2=\Lambda$ $\Rightarrow\;a(t)\propto e^{\sqrt{\Lambda/3}\,t}$, corresponds to a stage of inflationary ($q=-1$) de Sitter expansion. The eigenvalues of the matrix $J(P_{dS})$ are $\lambda_1=-3$ and $\lambda_2=-4$. This means that $P_{dS}$ is a future attractor (see the classification of isolated equilibrium points in the appendix \ref{appendix}). This entails that independent on the initial conditions chosen $\Omega_r^0$, $\Omega_\Lambda^0$, the orbits in $\Psi_\Lambda$ are always attracted towards the de Sitter state which explains the actual accelerated pace of the cosmic expansion in perfect fit with the observational data. This is why, in spite of the serious drawback in connection with the cosmological constant problem \cite{ccp, ccp-1}, the very simple $\Lambda$CDM model represents such a successful description of the present cosmological paradigm and it is, therefore, coined the ``concordance model''. Any other DE model, regardless of its nature, ought to be compared to the $\Lambda$CDM predictions as a first viability test.

Additional refinement of the above model is achieved if we add the energy density of baryons.

%------------------------------------------------------------------------

\section{Scalar field models of Dark Energy: the dynamical systems perspective}\label{scalar-f}

The cosmological constant problem can be split into two questions \cite{ccp}: (i) why the vacuum energy $\rho_{vac}=\Lambda$ is not very much larger? -- old cosmological constant problem -- and (ii) why it is of the same order of magnitude as the present mass density of the universe? which is acknowledged as the new cosmological constant problem. In order to avoid the old cosmological constant problem, which  is exclusive of the $\Lambda$CDM model, scalar field models of dark energy are invoked.\footnote{The new cosmological problem is inherent also in several scalar field models of dark energy.} In this last case an effective ``dynamical'' cosmological constant is described by the scalar field's $X$ self-interaction potential $V(X)$. This may be arranged in such a way that at early times the vacuum energy $\rho_{\text{vac},0}=V(X_0)$ ($\dot X=0$) is large enough as to produce the desired amount of inflation, while at late times $\rho_{\text{vac},f}=V(X_f)$ is of the same order of the CDM energy density $\rho_m$.

In this section, we shall explore the asymptotic structure of general scalar field models of DE \cite{copeland-rev, wands, coley, amendola, nunez, quiros-prd-2009, luis, lazkoz, boehmer, mimoso, barreiro, linder, de-models, dungan}. These represent a viable alternative to the $\Lambda$CDM model explored above. Among them ``exponential quintessence'' $V(X)=M\exp(-\mu X)$ \cite{wands, quintessence} is one of the most popular scalar field models of dark energy. 

The specification of a scalar field model for the DE means that the energy density and the parametric pressure in the cosmological field equations (\ref{x-cons-eq}) are given by

\bea\rho_X=\dot X^2/2+V(X),\;p_X=\dot X^2/2-V(X),\label{rho-p}\eea respectively. In these equations $V=V(X)$ is the self-interacting potential of the scalar field $X$. Besides, for the EOS parameter $\omega_X=\gamma_X-1$, one has 

\be \omega_X\equiv\frac{p_X}{\rho_X}=\frac{\dot X^2-2V}{\dot X^2+2V}.\label{q-eos}\ee After the above choice, the conservation equation (\ref{x-cons-eq}) can be written in the form of the following Klein-Gordon equation:

\bea \ddot X+3H\dot X=-V_{,X}.\label{kg-eq}\eea 

It happens that deriving an autonomous ODE for the variable $\omega_X$ -- see section \ref{model-indep} -- can be a very difficult task. Hence it could be better to choose a different set of phase space variables which do the job. Here, instead of the phase space variables $\Omega_X$ and $\omega_X$ we shall choose the following variables \cite{wands}:

\bea x\equiv\frac{\dot X}{\sqrt{6}H},\;y\equiv\frac{\sqrt V}{\sqrt{3}H},\label{xy-vars}\eea so that the dimensionless energy density of the scalar field $X$ is given by $\Omega_X=x^2+y^2$, and the Friedmann constraint (\ref{friedmann-c}) can be written as

\bea \Omega_m=1-x^2-y^2.\label{xy-friedmann-c}\eea

\subsection{Determination of the physical phase space}

The first step towards a complete study of the asymptotic structure of a given cosmological model is the rigorous determination of the phase space where to look for the relevant equilibrium points. In the present model there are several constraints on the physical parameters which help us to define the physically meaningful phase space. 

One of these constraints is given by (\ref{xy-friedmann-c}): $$0\leq\Omega_m\leq 1\;\Rightarrow\;0\leq x^2+y^2\leq 1.$$ Besides, since $0\leq\Omega_X\leq 1$, then $|x|\leq 1$. Additionally we shall be interested in cosmic expansion exclusively $H\geq 0$, so that $y\geq 0$. Hence, for instance, for potentials of one of the following kinds (see below): $V=V_0$, $V=V_0\,e^{\pm\mu X}$, the physical phase plane is defined as the upper semi-disk, 

\bea \Psi_X:=\{(x,y):0\leq x^2+y^2\leq 1,\,|x|\leq 1,\,y\geq 0\}.\label{psi-x}\eea However, as we shall see, in general -- arbitrary potentials -- the phase space is of dimension higher than 2 \cite{mimoso}.

%-------------------------------------

\begin{table*}\centering
\begin{tabular}{| c | c | c | c |}
\multicolumn{4}{c}{\textbf{Correspondence $f(z)\;\Leftrightarrow\;V(X)$}}\\
\hline \hline
Function $f(z)$ & $z$ & Self-interaction potential $V(X)$ &  Reference \\
\hline\hline
$f(z)=0$ & $\pm\mu$ & $V=V_0\,e^{\pm\mu X}$ & \cite{wands}\\
\hline
 $f(z)=-\alpha\beta+(\alpha+\beta)z-z^2$ & $\frac{\alpha+\beta\,e^{(\beta-\alpha)X}}{1+e^{(\beta-\alpha)X}}$ & $V=V_0\left(e^{\alpha X}+e^{\beta X}\right)$ & \cite{barreiro} \\
\hline 
$f(z)=(p^2\mu^2-z^2)/p$ & $p\mu\,\text{cotanh}(\mu X)$ & $V=V_0\sinh^p(\mu X)$ & \cite{sahni} \\
\hline 
$f(z)=(p^2\mu^2-z^2)/p$ & $p\mu\tanh(\mu X)$ & $V=V_0\cosh^p(\mu X)$ & \cite{sahni} \\
\hline
 $f(z)=(p^2\mu^2-z^2)/2p$ & $p\mu\frac{\sinh(\mu X)}{\cosh(\mu X)-1}$ & $V=V_0\left[\cosh(\mu X)-1\right]^p$ & \cite{wang}\\
\hline\hline
\end{tabular}\caption{The functions $f(z)$ in Eq. (\ref{f(z)-1}) for different self-interaction potentials.}\label{tab-fz}
\end{table*}

%------------------------------------

\subsection{Autonomous system of ODE-s}

In order to derive the autonomous ordinary equations for the phase space variables $x$, $y$ one proceeds in a similar fashion than in section \ref{model-indep}. First we writte the Raychaudhuri equation in (\ref{x-cons-eq}) 

\bea &&\dot H=-\frac{1}{2}\left(\gamma_m\rho_m+\rho_X+p_X\right)\;\Rightarrow\nonumber\\
&&-2\frac{\dot H}{H^2}=3\gamma_m\Omega_m+\frac{\dot X^2}{H^2},\nonumber\eea in terms of the new variables: 

\bea -2\frac{\dot H}{H^2}=3\gamma_m\left(1-x^2-y^2\right)+6x^2,\label{raycha-xy}\eea where we have taken into account the Friedmann constraint (\ref{xy-friedmann-c}). Then, given the definition (\ref{xy-vars}), let us find 

\bea \dot x=\frac{\ddot X}{\sqrt 6H}-\frac{\dot X}{\sqrt 6H}\frac{\dot H}{H},\label{der-1}\eea or if one substitutes (\ref{kg-eq}) and (\ref{raycha-xy}) back into (\ref{der-1}), then

\bea &&x'=-3x(1-x^2)+\nonumber\\
&&\;\;\;\;\;\;\;\;\;\frac{3\gamma_m}{2}\,x\left(1-x^2-y^2\right)-\sqrt\frac{3}{2}\frac{V_{,X}}{V}\,y^2,\label{x-ode}\eea where we replaced derivatives with respect to the cosmic time $t$ by derivatives with respect to the variable $\tau\equiv\ln a(t)$. Applying the same procedure with the variable $y$ one obtains $$y'=y\left(\frac{1}{2}\frac{V_{,X}}{V}\frac{\dot X}{H}-\frac{\dot H}{H^2}\right),$$ where we have taken into account that $\dot V=V_{,X}\dot X$, hence

\bea y'=y\left[\frac{3\gamma_m}{2}\left(1-x^2-y^2\right)+3x^2\right]+\sqrt\frac{3}{2}\frac{V_{,X}}{V}\,xy.\label{y-ode}\eea 

Unless $V=V_0$ $\Rightarrow\;V_{,X}/V=0$, or $V=V_0\,e^{\pm\mu X}$ $\Rightarrow\;V_{,X}/V=\pm\mu=const$., the ASODE (\ref{x-ode}), (\ref{y-ode}) is not a closed system of equations since one equation involving the derivative of $V_{,X}/V$ with respect to $\tau$ is lacking. 

One example where the ASODE (\ref{x-ode}), (\ref{y-ode}) is indeed a closed system of ODE-s is given by the so called ``exponential quintessence.'' This case has been investigated in all detail in \cite{wands} so that we submit the reader to that reference to look for a very interesting example of the application of the dynamical systems tools in cosmology. As a matter of fact, it results a very useful exercise for those who want to learn how to apply the dynamical systems tools in cosmological settings to reproduce the results of the dynamical systems study in the seminal work \cite{wands}.

In order to be able to consider self-interaction potentials beyond the constant and the exponential potentials, in addition to the variables $x$, $y$, one needs to adopt a new variable \cite{wang-1, z, nunez}

\bea z\equiv V_{,X}/V,\label{z-var}\eea so that $z=0$ corresponds to the constant potential, while $z=\pm\mu$ is for the exponential potential.\footnote{For non-exponential potentials, besides the increase in the dimensionality of the phase space, there is another problem. Typically the phase space becomes unbounded, so that two different sets of variables are required to cover the entire phase space \cite{mimoso}. In the present paper, however, the chosen example of the cosh-like potential does not lead to unbounded phase space. In consequence a single set of phase space variables is enough to describe the whole asymptotic (and intermediate) dynamics.} By taking the derivative of the new variable $z$ with respect to $\tau$ one obtains

\bea z'=\sqrt{6}\,xf(z),\label{z-ode}\eea where we have defined (as before take into account that $\dot V=V_{,X}\dot X$ and $\dot V_{,X}=V_{,XX}\dot X$, etc.)

\bea f(z)\equiv z^2\left[\Gamma-1\right],\;\Gamma\equiv VV_{,XX}/V^2_{,X},\label{f(z)}\eea and the main assumption has been that the above $\Gamma$ is a function of the variable $z$: $\Gamma=\Gamma(z)$. Notice that, since $$\Gamma\equiv\frac{VV_{,XX}}{V^2_{,X}}=\frac{V_{,XX}/V}{(V_{,X}/V)^2}=\frac{1}{z^2}\frac{V_{,XX}}{V},$$ the left-hand equation in (\ref{f(z)}) can be rewritten also as

\bea f(z)=\frac{V_{,XX}}{V}-z^2.\label{f(z)-1}\eea

In the event that $\Gamma$ can not be explicitly written as a function of $z$, then, an additional ODE: $\Gamma'=...$, is to be considered. However, this case is by far more complex and does not frequently arise.

Before going further we want to point out that, as a matter of fact, the function $\Gamma$ and the variable $z$ were first identified in \cite{wang-1}. Besides, the dynamical system (\ref{x-ode}), (\ref{y-ode}), (\ref{z-ode}) was explored in \cite{nunez} several years before it was studied in the reference \cite{z}. However, it was in \cite{z} where the possibility that for several specific self-interaction potentials $\Gamma$ can be explicitly written as a function of $z$, was explored for the first time. In the reference \cite{nunez}, although the correct dynamical system was identified, the authors were not interested in specific self-interaction potentials. The cost of the achieved generality of the analysis was that the authors had to rely on the obscure notion of ``instantaneous critical points''.

\section{An Example: The cosh-like potential}\label{cosh-sect}

In order to show with a concrete example how the function $f(z)$ can be obtained for an specific potential, let us choose the cosh potential \cite{sahni, wang} 

\bea V=V_0\cosh(\mu X).\label{cosh-pot}\eea This potential -- as well as a small variation of it in the last row of TAB. \ref{tab-fz} -- has a very interesting behavior near of the minimum at $X=0$. Actually, in the neighborhood of the minimum, the potential (\ref{cosh-pot}) approaches to $$V(X)\approx V_0+\frac{m^2}{2}X^2,\;m^2=V_0\mu^2.$$ The quadratic term is responsible for oscillations of the scalar field around the minimum which play the role of cold dark matter, hence, at late times when the system is going to stabilize in the minimum of $V(X)$ the cosh potential makes the scalar field model of DE to be indistinguishable from the standard $\Lambda$CDM model (see the properties of the equivalent model of Ref. \cite{quiros-prd-2009}). 

Taking derivatives of (\ref{cosh-pot}) with respect to the scalar field $X$ we have that $V_{,X}=\mu V_0\sinh(\mu X)$, and $V_{,XX}=\mu^2V$. Hence, 

\bea z=V_{,X}/V=\mu\tanh(\mu X),\;f(z)=\mu^2-z^2.\label{cosh-fz}\eea Notice that the $z$-s which solve $f(z)=0$, i. e., $z=\pm\mu$, which correspond to the exponential potentials, are critical points of (\ref{z-ode}). In TAB. \ref{tab-fz} the functions $f(z)$ for several well-known potentials are displayed.

Let us to collect all of the already found autonomous ODE-s which correspond to a scalar field model of DE with arbitrary self-interaction potential:

\bea &&x'=-3x(1-x^2)+\nonumber\\
&&\;\;\;\;\;\;\;\;\;\frac{3\gamma_m}{2}\,x\left(1-x^2-y^2\right)-\sqrt\frac{3}{2}\,y^2z,\nonumber\\
&&y'=y\left[\frac{3\gamma_m}{2}\left(1-x^2-y^2\right)+3x^2\right]+\sqrt\frac{3}{2}\,xyz,\nonumber\\
&&z'=\sqrt{6}\,xf(z),\label{master-ode}\eea where it is evident that the equations for $x$ and $y$ are independent of the self-interaction potential, and that details of the given model are encoded in the function $f(z)$ -- Eq. (\ref{f(z)-1}) -- which depends on the concrete form of the potential (see TAB. \ref{tab-fz}).

%-------------------------------------

\begin{table*}\centering
\begin{tabular}{|c|c|c|c|c|c|c|c|c|c|}
\hline\hline
Critical Points&\;\;\;$x$\;\;\;&\;\;\;$y$\;\;\;&\;\;\;$z$\;\;\;&Existence&\;\;\;$\Omega_X$\;\;\;&\;\;\;$\Omega_m$\;\;\;&\;\;\;$\lambda_1$\;\;\;&\;\;\;$\lambda_2$\;\;\;&\;\;\;$\lambda_3$\;\;\;\\
\hline\hline
${\cal Z}$ & $0$ & $0$ & $z$ & Always & $0$ & $1$ & $3/2$ & $-3/2$ & $0$ \\
\hline
$P_{dS}$ & $0$ & $1$ & $0$ & '' & $1$ & $0$ & $-\frac{3}{2}+\frac{3}{2}\sqrt{1-\frac{4\mu^2}{3}}$ & $-\frac{3}{2}-\frac{3}{2}\sqrt{1-\frac{4\mu^2}{3}}$ & $-3$ \\
\hline
$P^\pm_K$ & $\pm 1$ & $0$ & $\mu$ & '' & $1$ & $0$ & $3\pm\sqrt\frac{3}{2}\mu$ & $\mp 2\sqrt{6}\mu$ & $3$ \\
\hline
$\bar P^\pm_K$ & $\pm 1$ & $0$ & $-\mu$ & '' & $1$ & $0$ & $3\mp\sqrt\frac{3}{2}\mu$ & $\pm 2\sqrt{6}\mu$ & $3$ \\
\hline
$P^\pm_{X/m}$ & $\pm\sqrt\frac{3}{2}\frac{1}{\mu}$ & $\sqrt\frac{3}{2}\frac{1}{\mu}$ & $\mp\mu$ & $\mu^2\geq\frac{3}{2}$ & $\frac{3}{\mu^2}$ & $1-\frac{3}{\mu^2}$ & $-\frac{3}{4}+\frac{3}{4}\sqrt{\frac{24}{\mu^2}-7}$ & $-\frac{3}{4}-\frac{3}{4}\sqrt{\frac{24}{\mu^2}-7}$ & $6$ \\
\hline
$P^\pm_X$ & $\pm\frac{\mu}{\sqrt 6}$ & $\sqrt{1-\frac{\mu^2}{6}}$ & $\mp\mu$ & $\mu^2\leq 6$ & $1$ & $0$ & $-3+\mu^2$ & $-3+\frac{\mu^2}{2}$ & $2\mu^2$ \\
\hline\hline
\end{tabular}\caption{Critical points of the autonomous system of ODE (\ref{cosh-ode}) and their properties.}\label{tab-cosh}
\end{table*}
%----------------------------------------

\subsection{Critical points}

If we apply the simplest tools of the dynamical systems theory to the the above example with the cosh potential the corresponding asymptotic dynamics in the phase space is revealed. Given that, at the minimum of the cosh potential, the model behaves as $\Lambda$CDM, here we shall assume that the matter fluid is composed mainly of baryons so that it behaves like pressureless dust ($\gamma_m=1$). 

The following closed system of autonomous ODE-s is obtained (it is just Eq. (\ref{master-ode}) with the appropriate substitutions):

\bea &&x'=-\frac{3}{2}\,x\left(1-x^2+y^2\right)-\sqrt\frac{3}{2}\,y^2z,\nonumber\\
&&y'=\frac{3}{2}\,y\left(1+x^2-y^2\right)+\sqrt\frac{3}{2}\,xyz,\nonumber\\
&&z'=\sqrt{6}\,x (\mu^2-z^2).\label{cosh-ode}\eea Since $z=\mu\tanh(\mu X)$ $\Rightarrow\;-\mu\leq z\leq\mu,$ the phase space where to look for critical points of (\ref{cosh-ode}) is the bounded semi-cylinder (see Eq. (\ref{psi-x})):

\bea &&\Psi_\text{cosh}:=\{(x,y,z):0\leq x^2+y^2\leq 1,\nonumber\\
&&\;\;\;\;\;\;\;\;\;\;\;\;\;\;\;\;\;\;\;\;\;\;\;|x|\leq 1,\,y\geq 0,|z|<\mu\}.\label{psi-3d}\eea 

Nine critical points $P_i:(x_i,y_i,z_i)$ of the autonomous system of ODE (\ref{cosh-ode}) and one critical manifold are found in $\Psi_\text{cosh}$. These together with their main properties are shown in TAB. \ref{tab-cosh}. For sake of conciseness, here we shall concentrate only in a few of them which have some particular interest.

The de Sitter critical point $P_{dS}:(0,1,0)$, for which $\Omega_X=1$ ($\Omega_m=0$), is associated with accelerating expansion since the deceleration parameter 

\bea q=-1-\frac{\dot H}{H^2}=\frac{1}{2}+\frac{3}{2}\left(x^2-y^2\right)=-1,\label{q}\eea is a negative quantity. Since $z=0\;\Rightarrow\;V=V_0$, and since $x=0\;\Rightarrow\;\dot X=0$, and $$y=1\;\Rightarrow\;H=\sqrt\frac{V_0}{3}\;\Rightarrow a(t)\propto e^{Ht},$$ this point corresponds to a de Sitter (inflationary) phase of the cosmic expansion. We underline the fact that to a given point in the phase space -- in this case $P_{dS}\in\Psi_\text{cosh}$ -- it corresponds a specific cosmological dynamics (de Sitter expansion in the present case).

Next we have the scalar field $X$-dominated equilibrium points 

\bea &&P^\pm_X:\left(\pm\frac{\mu}{\sqrt 6},\sqrt{1-\frac{\mu^2}{6}},\mp\mu\right),\;\Omega_X=1,\;q=-1+\frac{\mu^2}{2}.\nonumber\eea The first thing one has to care about is the existence of the critical point. By existence we mean that the given point actually belongs in the phase space ($\Psi_\text{cosh}$ in the present case). Hence, since $|x|\leq 1\;\Rightarrow\;\mu^2\leq 6$. This is precisely the condition for existence of the points $P^\pm_X$: $\mu^2\leq 6$. This warrants, besides, that $y=\sqrt{1-\mu^2/6}$ be a real number. If one substitutes (\ref{xy-vars}) back into (\ref{q-eos}) on gets $$\omega_X=\frac{x^2-y^2}{x^2+y^2},$$ so that, for the present case the EOS parameter of the $X$-field $\omega_X=\mu^2/3-1\;\Rightarrow\;\gamma_X=\mu^2/3$. Then the continuity equation (\ref{x-cons-eq}) can be readily integrated at $P^\pm_X$ to obtain: $\rho_X=M\,a^{-3\gamma_X}=M\,a^{-\mu^2}$, where $M$ is an integration constant. After this the Friedmann constraint $$\Omega_X=1\;\Rightarrow\;3H^2=\rho_X\;\Rightarrow\;\frac{\dot a}{a}=\sqrt\frac{M}{3}\,a^{-\mu^2},$$ can be integrated also to obtain the following cosmological dynamics $a(t)\propto t^{2/\mu^2}$, which is to be associated with the critical points $P^\pm_X$. We want to point out that whenever $2<\mu^2\leq 6$ ($q>0$) the present scalar field dominated solution represents decelerated expansion.

In the case of the matter-dominated critical manifold\footnote{Here by critical manifold we understand a curve in the phase space all of whose points are critical points.} $${\cal M}=\{(0,0,z):z\in R\},\;\Omega_m=1,\;q=1/2,$$ instead of an isolated equilibrium point this is actually a 1-dimensional manifold which is extended along the $z$-direction. This means in turn that the matter-dominated solution: $\Omega_m=1\;\Rightarrow$ $$3H^2=\rho_m=M a^{-3}\;\Rightarrow\;a(t)\propto t^{2/3},$$ may coexist with the scalar field domination no matter whether $V\propto\cosh(\mu X)$ $\rightarrow$ intermediate $X$-s, $V=V_0$ $\rightarrow$ $X=0$, or $V\propto e^{\pm\mu X}$ $\rightarrow$ very large $X$-s ($|X|\rightarrow\infty$). This entails that it can be found an heteroclinic orbit in the phase portrait -- an orbit connecting two or more different equilibrium points -- joining the matter-dominated equilibrium state with a scalar field domination critical point, be it originated either by a cosmological constant, or by exponential or cosh-like potentials.

\subsection{Stability}

In order to judge about the stability of given hyperbolic equilibrium points (as explained in the appendix \ref{appendix}) one needs to linearize the system of ODE -- in this case (\ref{cosh-ode}) -- which means, in the end, that we have to find the eigenvalues of the linearization or Jacobian matrix $J$ (see Eq. (\ref{matrix-eq}) of the appendix) $$J=\begin{pmatrix} \der x'/\der x & \der x'/\der y & \der x'/\der z \\ \der y'/\der x & \der y'/\der y & \der y'/\der z \\ \der z'/\der x & \der z'/\der y & \der z'/\der z \end{pmatrix}.$$ Next $J$ is to be evaluated at each non-hyperbolic critical point, for instance, $J(P_{dS})$, $J(P_X)$, etc. The eigenvalues of the resulting numerical matrices is what we use to judge about the stability of the given points. 

Take, for instance, the de Sitter point $P_{dS}:(0,1,0)$. In this case one has to find the roots of the following algebraic equation for the unknown $\lambda$: $$\det\begin{pmatrix} -3-\lambda & 0 & \sqrt{3/2}\mu\\ 0 & -3-\lambda & 0 \\ -\sqrt{6}\mu & 0 & -\lambda \end{pmatrix}=0.$$ The resulting eigenvalues are $$\lambda_{1,2}=-\frac{3}{2}\pm\frac{3}{2}\sqrt{1-\frac{4}{3}\,\mu^2},\;\lambda_3=-3.$$ For $\mu^2\leq 3/4$ the critical point is an isolated focus, while for $\mu^2>3/4$, since the eigenvalues $\lambda_1$ and $\lambda_2$ are complex numbers with negative real parts, the point $P_{dS}$ is a stable spiral. Hence the de Sitter solution is always a future attractor (see the bottom panel of FIG. \ref{fig-lcdm}). This is consistent with the fact that the cosh potential is a minimum at $X=0\;\Rightarrow\;V=V_0$.

The eigenvalues of the linearization matrix $J(P^\pm_X)$ are $$\lambda_1=-3+\mu^2/2,\;\lambda_2=-3+\mu^2,\;\lambda_3=\mu^2.$$ Hence, since as required by the existence $\mu^2<6$, $P^\pm_X$ are always saddle critical points (at least one of the eigenvalues is of a different sign). 

If one computes the eigenvalues of $J({\cal M})$ one finds $\lambda_{1,2}=\pm 3/2,\;\lambda_3=0.$ The vanishing eigenvalue is due to the fact that the 1-dimensional manifold extends along the $z$-direction.\footnote{In general a critical point for which at least one of the real parts of the eigenvalues of the Jacobian matrix is vanishing, is called as a non-hyperbolic critical point. In such case the Hartman-Grobman theorem \cite{hartman-grobman} can not be applied. As a consequence one has to go beyond the linear stability theory. The centre manifold theory is useful in this case. For the application of the centre manifold theory in a case of cosmological interest see \cite{lazkoz, boehmer}.} The differing signs of the eigenvalues $\lambda_1$ and $\lambda_2$ means that each point in ${\cal M}$ is a saddle. It might be a nice task for the interested readers to judge about the stability of the remaining critical points of (\ref{cosh-ode}) (see TAB. \ref{tab-cosh}).

%----------------------------------------

\section{Conclusion}\label{conclu}

In this paper we have shown in as much as possible pedagogical way, how the application of the basic tools of the dynamical systems theory can help us to deeper understand the dynamics of cosmological models of dark energy. Our main aim has been to make clear how much useful information about the asymptotic cosmological dynamics -- the one which decides the origin and the fate of our Universe -- can be extracted by means of these simple tools. We have concentrated our exposition in scalar field cosmological models of dark energy and the analysis was kept as general as possible. Special attention has been paid to the derivation of relevant equations and formulas, so as to facilitate the way for beginners to concrete computations. 

It has been demonstrated that certain generic asymptotic behavior arises which is quite independent on the concrete model of DE: matter dominance transient phase and dark energy dominance at late times. Just for illustration we have chosen a concrete potential -- the cosh potential -- which at late times approaches to the $\Lambda$CDM model. The approach undertaken here can be easily applied to other self-interaction potentials (see TAB. \ref{tab-fz}).

In the present paper, for completeness, we have included a very concise and simple exposition of the fundamentals of the theory of the dynamical systems -- see the appendix \ref{appendix} --  which may be considered as material for an introductory course on this subject.

\section*{ACKNOWLEDGMENT}

The authors thank Fernando R. Gonz\'alez-D\'iaz and Ricardo Medel-Esquivel for useful comments on the original version of the manuscript and the SNI of Mexico for support. The work of R G-S was partly supported by SIP20131811, SIP20140318, COFAA-IPN, and EDI-IPN grants. The research of T G, F A H-R, and I Q was partially supported by Fondo Mixto 2012-03, GTO-2012-C03-194941. I Q was supported also by CONACyT of Mexico. Last but not least the authors are indebted to the anonymous referees for the constructive criticism.

%---------------------------------------------------

\section{Appendix: dynamical systems in cosmology}\label{appendix}

In classical mechanics the state of a given physical system composed, say, of $N$ particles, is completely specified by the knowledge of the $N$ generalized coordinates $q_i$ ($i=1,...,N$) and the $N$ conjugated momenta $p_i$, which satisfy the Hamilton's canonical equations of motion $$\dot q_i=\partial{\cal H}/\partial p_i,\;\dot p_i=-\partial{\cal H}/\partial q_i,$$ where ${\cal H}={\cal H}(q_1,...,q_N,p_1,...,p_N,t)$ is the Hamiltonian of the physical system. Hence, a given state of this system corresponds to a point in the $2N$- dimensional space of states spanned by the $2N$ coordinates $q_1,...,q_N,p_1,...,p_N,$ also known as the ``phase space.''

Like it is for any other physical system, the possible states of a cosmological system may be also correlated with the points in an equivalent state space. However, unlike in the classical mechanics case, the choice of the variables -- generalized coordinates and their conjugate momenta -- for a cosmological model is not a trivial issue. In this case a certain degree of uncertainty in the choice of an appropriate set of variables of the phase space arises. There are, however, certain not written rules one follows when choosing appropriate variables of the phase space: (i) the variables should be dimensionless, and (ii) whenever possible, these should be bounded. The latter requirement is necessary to have a bounded phase space where all of the existing equilibrium points are ``visible,'' i. e., none of then goes to infinity.

The interplay between a cosmological model and the corresponding phase space is possible due to an existing one-to-one correspondence between exact solutions of the cosmological field equations (\ref{x-cons-eq}) and points in the phase space spanned by given variables $x$, $y$:\footnote{Here, for simplicity of the exposition, we consider a 2D phase space.} $$x=x(H,\rho_x,\rho_X),\;y=y(H,\rho_x,\rho_X).$$ When we replace the original field variables $H$, $\rho_x$, and $\rho_X$, by the phase space variables $x$, $y$, we have to keep in mind that, at the same time, we trade the original set of non-linear second order differential equations in respect to the cosmological time $t$ (equations (\ref{x-cons-eq})), by a set of first order ordinary differential equations (ODE): 

\bea x'=f(x,y),\;y'=g(x,y),\label{asode}\eea where the tilde denotes derivative with respect to the dimensionless ``time'' parameter, $\tau\equiv\int da/a\;\Rightarrow\;d\tau=d\ln a=H dt$. The introduction of $\tau$ instead of the cosmic time $t$ is dictated by simplicity and compactness of writing. Besides, this warrants that the first order differential equations (\ref{asode}) are a closed system of equations.

The most important feature of the system of ODE (\ref{asode}) is that the functions $f(x,y)$ and $g(x,y)$ do not depend explicitly on the parameter $\tau$. This is why (\ref{asode}) is called as an autonomous system of ODE. The image of the integral curves of (\ref{asode}) in the phase space are called ``orbits'' of the system of autonomous ODE. The critical points of (\ref{asode}) -- also fixed or stationary points -- $P_{cr}:(x_{cr},y_{cr})$ are those for which the ``velocity'' vector ${\bf v}=(x',y')$ vanishes (see below): $${\bf v}(P_{cr})=(x'(P_{cr}),y'(P_{cr}))=0.$$ 

In order to judge about the stability of given equilibrium points one needs to linearly expand (\ref{asode}) in the neighborhood of each hyperbolic critical point: $$f(x_{cr}+\delta x,y_{cr}+\delta y),\;g(x_{cr}+\delta x,y_{cr}+\delta y).$$ Depending on whether the linear perturbations $\delta x=\delta x(\tau)$, $\delta y=\delta y(\tau)$ decay (grow) with time $\tau$ or decay in one direction while grow in the other, the critical point can be a future (past) attractor or a saddle point. 

Knowledge of the critical -- also equilibrium or fixed -- points in the phase space corresponding to a given cosmological model is a very important information. In particular, the existence of attractors can be correlated with generic cosmological solutions that might really decide the fate and/or the origin of the cosmic evolution. The ``meta-stable'' saddle equilibrium points -- stationary points which are not local extrema -- can be associated with (not less important) transient cosmological solutions. Hence, in the end we are trading the study of the cosmological dynamics depicted by $H=H(t)$, $\rho_x=\rho_x(t)$, $\rho_X=\rho_X(t)$, by the study of the properties of the equilibrium points of the equivalent autonomous system of ODE (\ref{asode}) in the phase space $\Psi$ of the cosmological model, which, we recall, differs from the usual definition in classical mechanics.

%---------------------------------------------------

\begin{figure*}[t!]\begin{center}
\includegraphics[width=4cm]{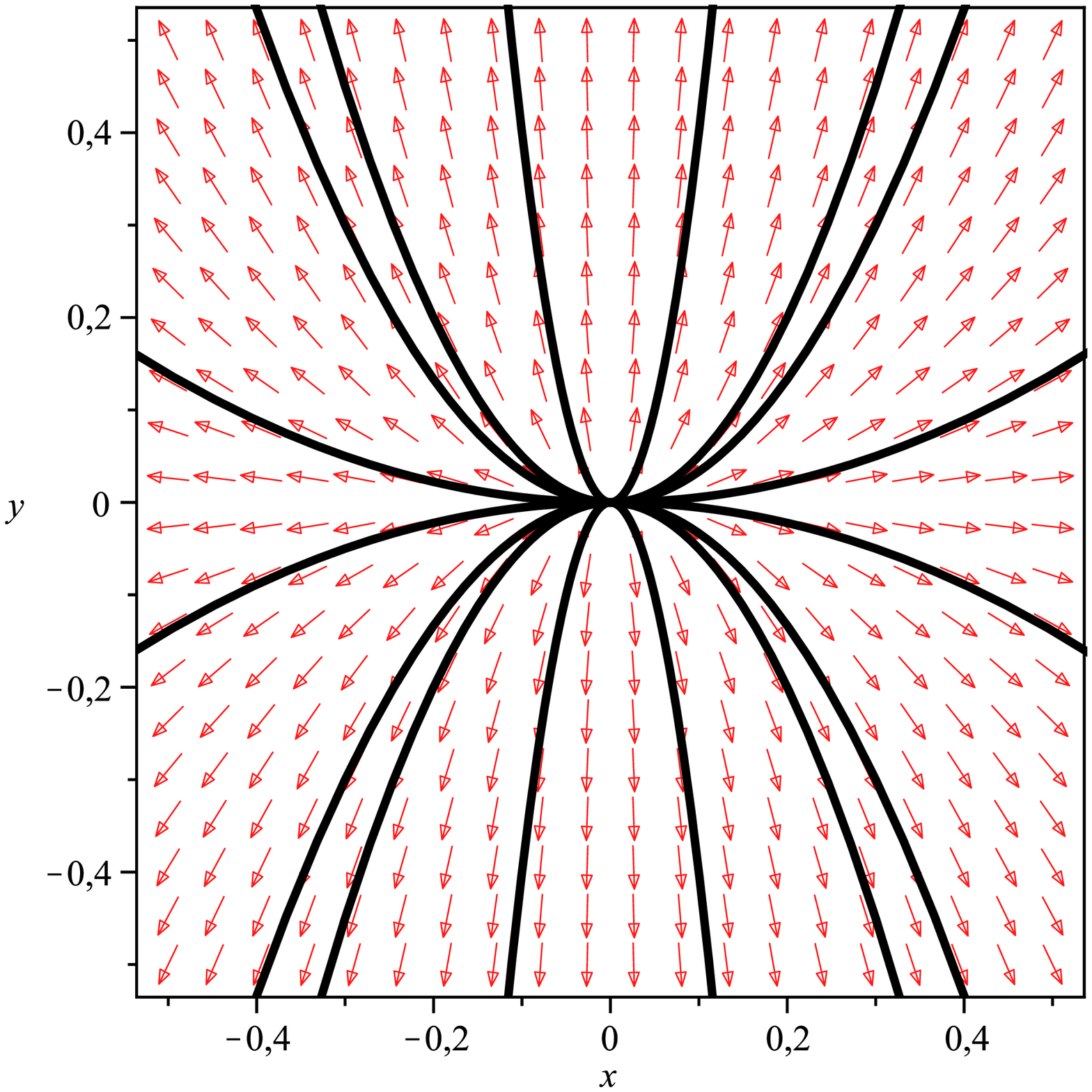}
\includegraphics[width=4cm]{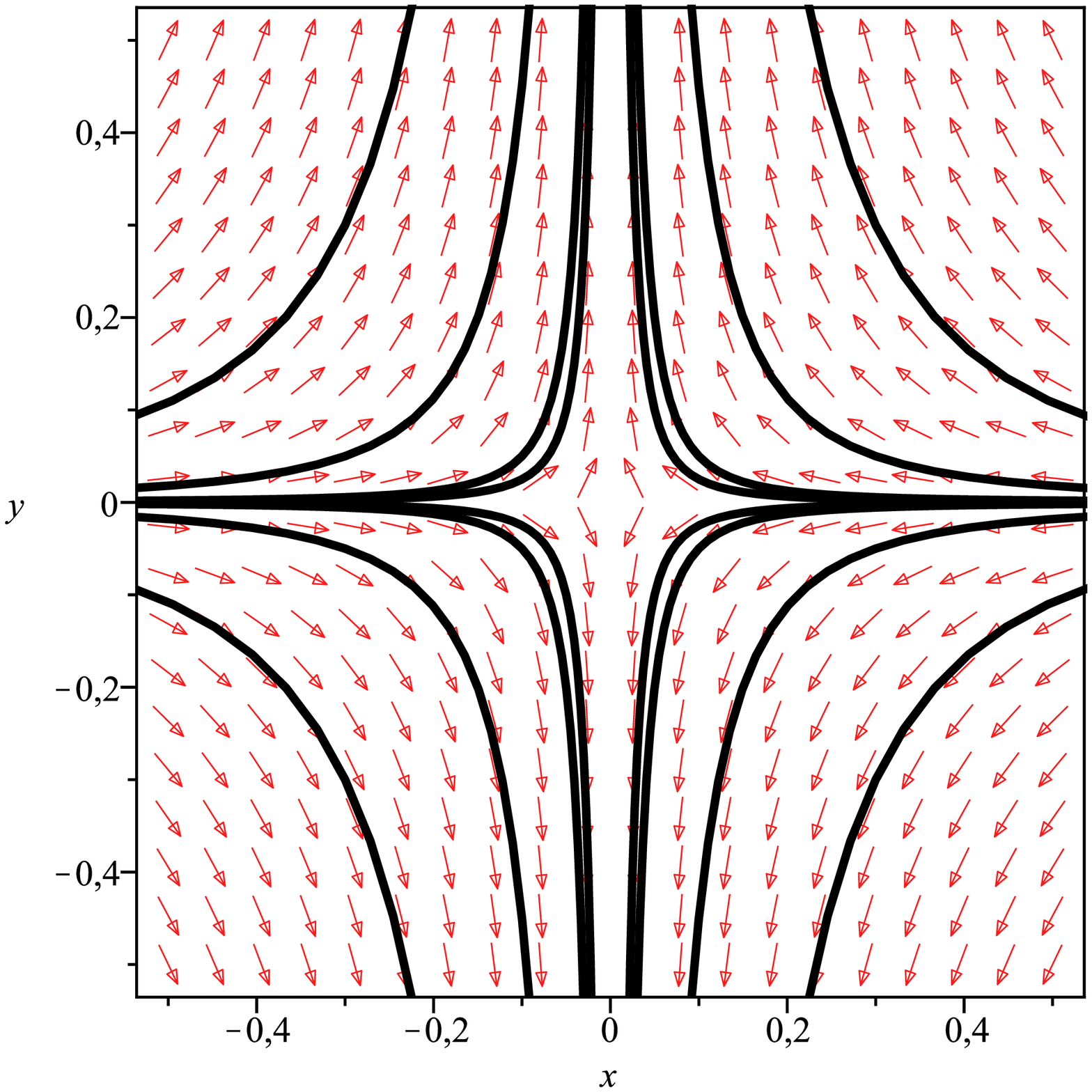}
\includegraphics[width=4cm]{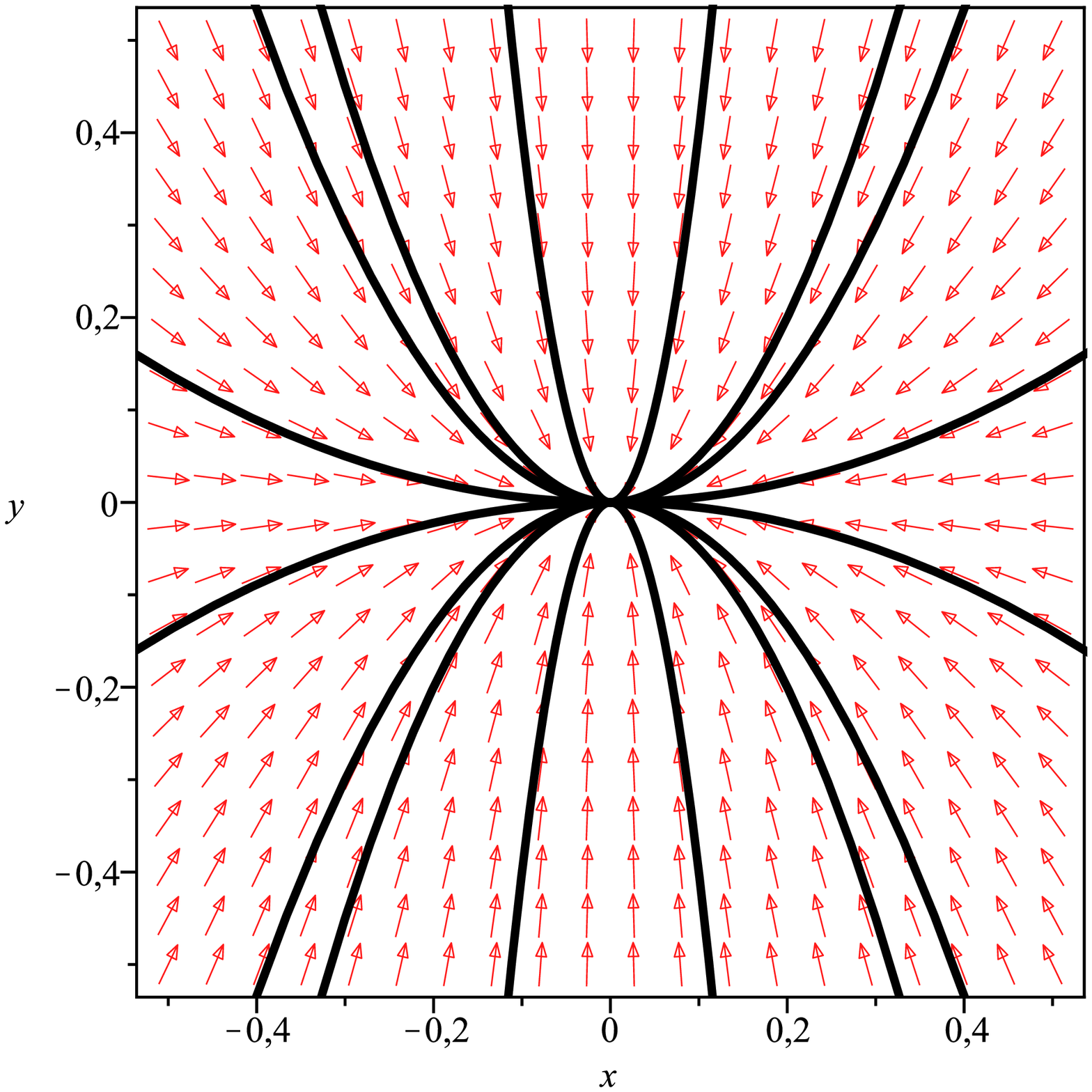}\\
\includegraphics[width=4cm]{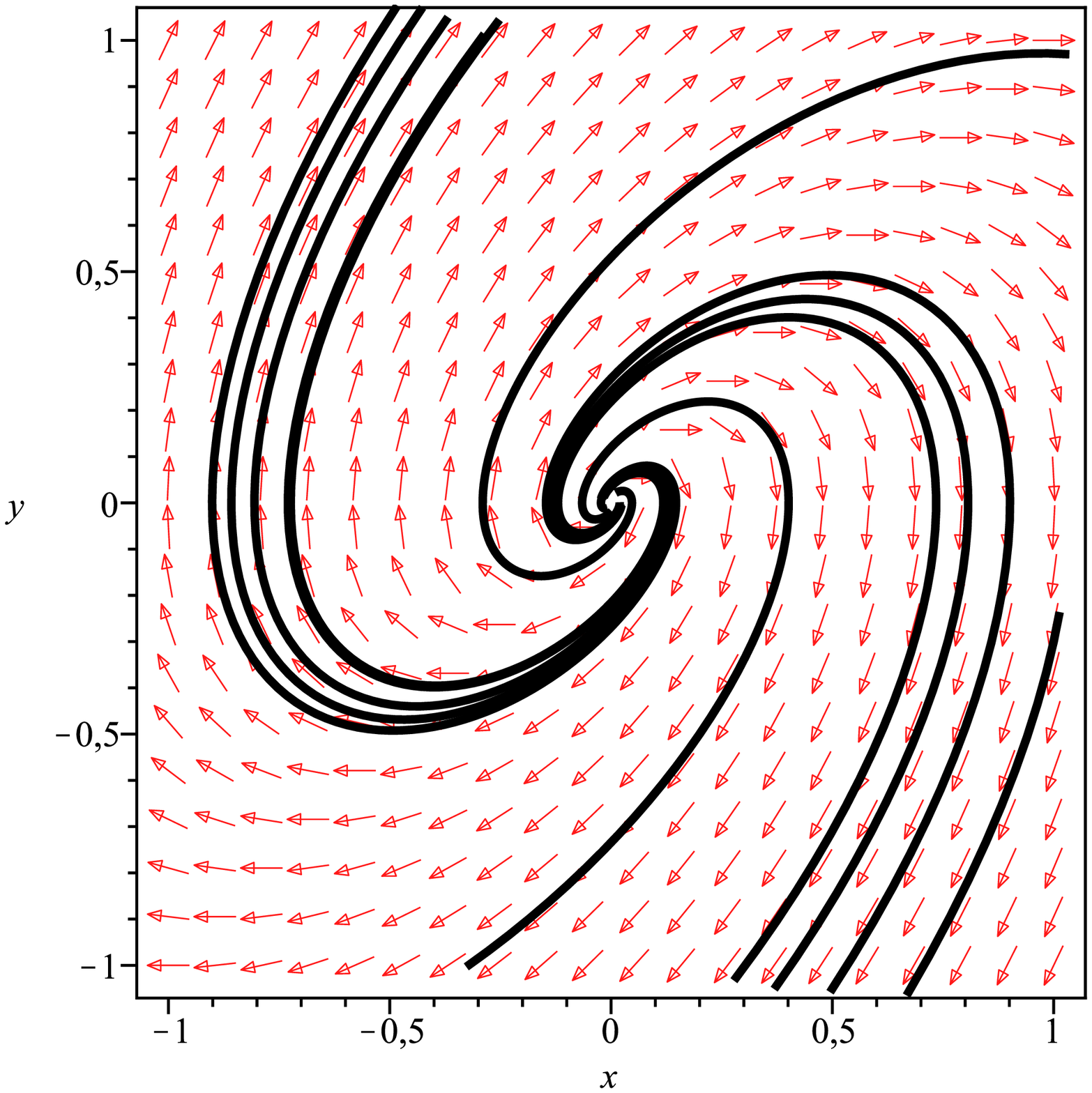}
\includegraphics[width=4cm]{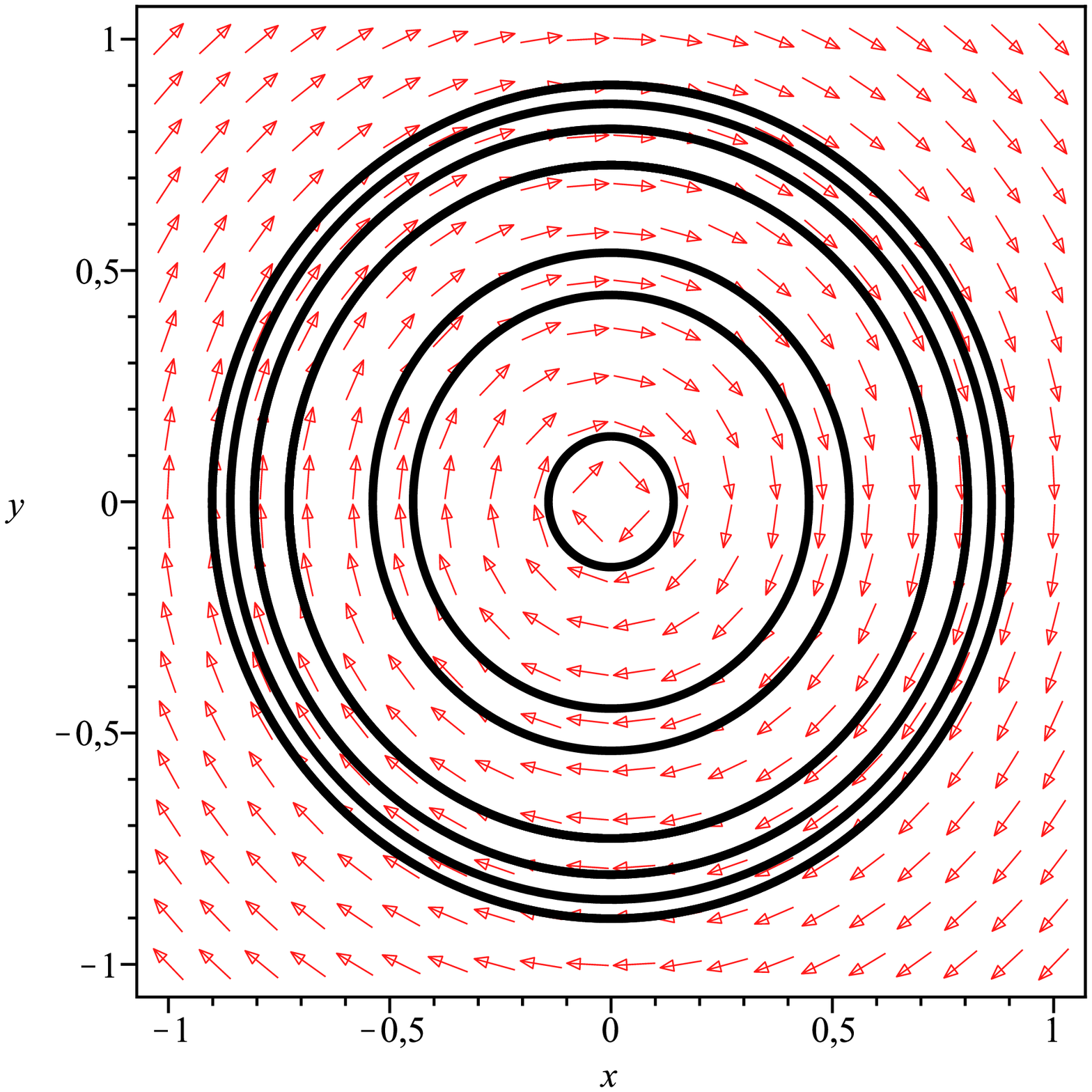}
\includegraphics[width=4cm]{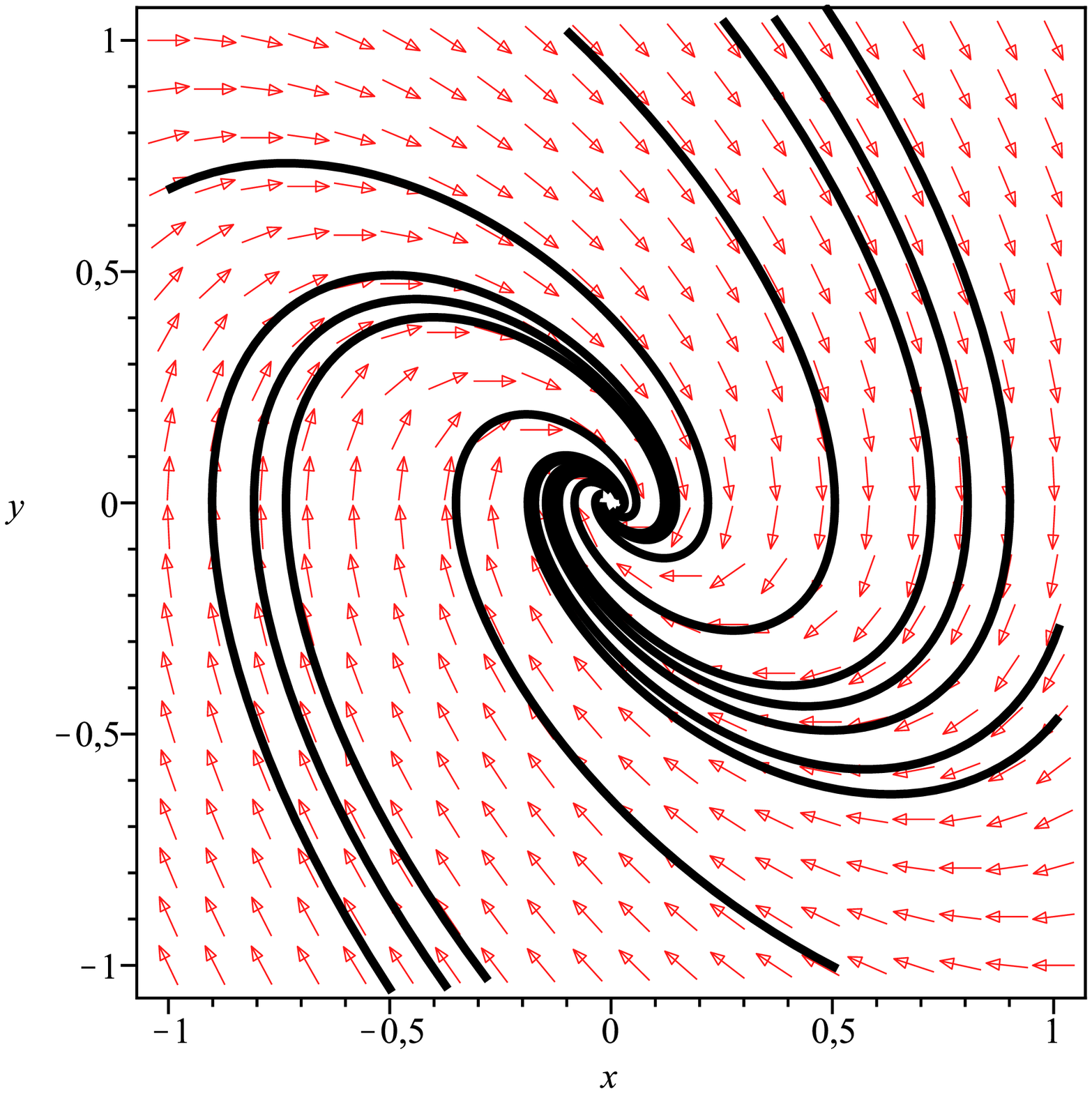}\vspace{0.7cm}
\caption{Phase portrait: Isolated equilibrium points in the phase space when the eigenvalues $\lambda_1$ and $\lambda_2$ are real numbers (top panels). The velocity field, represented by the arrows field in the figures, shows where the flux of the ODE (\ref{asode}) flows. Several orbits generated by different sets of initial conditions $[x(0)=x_{0i};y(0)=y_{0i}]$ are shown in each case. In the left-hand top panel the origin $P:(0;0)$ is a unstable node or past attractor ($\lambda_2>\lambda_1>0$), while in the center top panel the origin is a saddle critical point ($\lambda_1<0<\lambda_2$), and in the right-hand top panel $P:(0;0)$ is a stable node or future attractor ($\lambda_1<\lambda_2<0$). In the bottom panels the equilibrium points for the case when the eigenvalues are complex numbers [$\lambda_{1,2}=\nu\pm i\omega$] are shown. In the left-hand bottom panel the origin $(x;y)=(0;0)$ is a unstable spiral ($\nu>0$), while, in the center panel it is a center ($\nu=0$). The point $(0;0)$ is a stable spiral in the right-hand bottom panel ($\nu<0$).}\label{fig-clas-p-crit}\end{center}
\end{figure*}

%------------------------------------------------

\subsection{Remarks on phase space analysis}

In this subsection we provide a very simplified exposition of the fundamentals of the dynamical systems theory which are useful in most cosmological applications \cite{wands, coley, copeland-rev}. Only knowledge of elementary linear algebra and of ordinary differential equations is required to understand the material exposed here. For a more formal and complete introduction to this subject we recommend the well-known texts \cite{hirsh, brauer, arrowsmith, perko, arnold, ellis}.

The critical points of the system of ODE (\ref{asode}) $P_i:(x_i;y_i)$, i. e., the roots of the system of algebraic equations $$f(x,y)=0,\;g(x,y)=0,$$ correspond to privileged or generic solutions of the original system of cosmological equations. To judge about their stability properties it is necessary first to linearize (\ref{asode}) around the hyperbolic equilibrium points.\footnote{$P_h$ is an hyperbolic critical point if the real parts of all of the eigenvalues of the linearization matrix around $P_h$ are necessarily non-vanishing. In particular, an hyperbolic point can not be a center. In more technical words: an hyperbolic critical point is a fixed point that does not have any center manifolds.} To linearize around a given hyperbolic equilibrium point $P_i$ amounts to consider small linear perturbations $$x\rightarrow x_i+\delta x(\tau),\;y\rightarrow y_i+\delta y(\tau).$$ These perturbations would obey the following system of coupled ODE (here we use matrix notation):

\bea \delta{\bf x}'=J(P_i)\cdot\delta{\bf x},\;\delta{\bf x}=\begin{pmatrix}\delta x\\\delta y\end{pmatrix},\;J=\begin{pmatrix}\frac{\der f}{\der x}&\frac{\der f}{\der y}\\\frac{\der g}{\der x}&\frac{\der g}{\der y}\end{pmatrix}.\label{matrix-eq}\eea where $J$ - the Jacobian (also linearization) matrix, is to be evaluated at $P_i$. Thanks to the Hartman-Grobman theorem \cite{hartman-grobman}, which basically states that the behavior of a dynamical system in the neighborhood of each hyperbolic equilibrium point is qualitatively the same as the behavior of its linearization, we can safely replace the study of the dynamics of (\ref{asode}) by the corresponding study of its linearization (\ref{matrix-eq}).  

We assume that $J$ can be diagonalized, i. e., $J_D=M^{-1} J M$, where $M$ is the diagonalization matrix and $$J_D=\begin{pmatrix}\lambda_1&0\\0&\lambda_2\end{pmatrix},$$ is the diagonal matrix whose non vanishing components are the eigenvalues of the Jacobian matrix $J$:

\bea \det|J-\lambda U|=0,\;U=\begin{pmatrix}1&0\\0&1\end{pmatrix}.\label{caract-ec}\eea

After diagonalization the coupled system of ODE (\ref{matrix-eq}) gets decoupled:

\bea \delta\bar{\bf x}'={\it J_D}\cdot\delta{\bar{\bf x}},\;\delta{\bar{\bf x}}={\it M}^{-1}\delta{\bf x},\label{decop-edo}\eea where we have to recall that to each equilibrium point it corresponds a different matrix $J_D$. The decoupled system of ODE (\ref{decop-edo}) is easily integrated: $$\delta\bar x(\tau)=\delta\bar x(0)\,e^{\lambda_1\tau},\;\delta\bar y(\tau)=\delta\bar y(0)\,e^{\lambda_2\tau}.$$ Since the diagonal perturbations $\delta\bar x$ and $\delta\bar y$ are linear combinations of the perturbations $\delta x$, $\delta y$: $$\delta\bar x=c_{11}\delta x+c_{12}\delta y,\;\delta\bar y=c_{21}\delta x+c_{22}\delta y,$$ where the constants $c_{ij}$ are the components of the matrix $M^{-1}$, then  

\bea &&\delta x(\tau)=\bar c_{11}\,e^{\lambda_1\tau}+\bar c_{12}\,e^{\lambda_2\tau},\nonumber\\
&&\delta y(\tau)=\bar c_{21}\,e^{\lambda_1\tau}+\bar c_{22}\,e^{\lambda_2\tau},\label{perts}\eea where

\bea &&\bar c_{11}=\frac{c_{22}\delta\bar x(0)}{c_{22}c_{11}-c_{12}c_{21}},\;\bar c_{12}=-\frac{c_{12}\delta\bar y(0)}{c_{22}c_{11}-c_{12}c_{21}},\nonumber\\
&&\bar c_{21}=-\frac{c_{21}\delta\bar x(0)}{c_{22}c_{11}-c_{12}c_{21}},\;\bar c_{22}=\frac{c_{11}\delta\bar y(0)}{c_{22}c_{11}-c_{12}c_{21}}.\nonumber\eea 

As a matter of fact we do not need to compute the coefficients $\bar c_{ij}$, instead the structure of the eigenvalues $\lambda_i$ is the only thing we need to judge about the stability of given (hyperbolic) equilibrium points of (\ref{asode}). If the eigenvalues are complex numbers $\lambda_\pm=\nu\pm i\omega$ the perturbations (\ref{perts}) do oscillations with frequency $\omega$. If the real part $\nu$ is positive the oscillations are enhanced, while if $\nu<0$ the oscillations are damped. The case $\nu=0$ is associated with a center -- harmonic oscillations -- and is not frequently encountered in cosmological applications. This latter kind of point in the phase space -- eigenvalue with vanishing real part -- is called as non-hyperbolic critical point.

\subsection{Taxonomy of isolated equilibrium points} 

Here we summarize the basic classification of isolated equilibrium points in the phase plane. For simplicity, we assume that $\lambda_1\leq\lambda_2$, but nothing changes if one assumes that $\lambda_1\geq\lambda_2$, or if one makes no assumption at all. In the figure \ref{fig-clas-p-crit} the illustrative phase portraits -- the drawing of the trajectories of a dynamical system in the phase plane -- are shown.

\begin{enumerate}

\item The eigenvalues $\lambda_1$, $\lambda_2$ in Eq. (\ref{perts}) are real numbers.
    
		\begin{enumerate}
		
		\item $0<\lambda_1<\lambda_2$ -- the critical point $P_i$ is an unstable node or, also, a source point (past attractor). This is a unstable critical point which represents the origin of a non-empty set of orbits in the phase plane (left-hand top panel of FIG. \ref{fig-clas-p-crit}).
				
		\item $\lambda_1<0<\lambda_2$ -- $P_i$ is a saddle point. This is a unstable equilibrium point but can be associated with a marginally stable state since the orbits of the system of autonomous ODE-s (\ref{asode}) spend some ``time'' in the neighborhood $N(\epsilon,P_i)$ of the critical point until these leave $N(\epsilon,P_i)$ to go elsewhere in the phase plane (center top panel of FIG. \ref{fig-clas-p-crit}).
		
		\item $\lambda_1<\lambda_2<0$ (also $\lambda_1=\lambda_2<0$) -- $P_i$ is a stable node (future attractor). This kind of equilibrium point is associated with an asymptotically stable state. The future attractor is the end point of a non-empty set of phase space orbits generated by a wide range of initial conditions (right-hand top panel of FIG. \ref{fig-clas-p-crit}).
		
		\end{enumerate}

\item The eigenvalues $\lambda_1$, $\lambda_2$ in Eq. (\ref{perts}) are complex numbers: $\lambda_{1,2}=\nu\pm i\omega$ ($\omega\neq 0$). 

    \begin{enumerate}
		
		\item $\nu>0$ -- the equilibrium point $P_i$ is an unstable spiral (past attractor). It is associated with unstable oscillations, i. e., with amplified oscillations (left-hand bottom panel in FIG. \ref{fig-clas-p-crit}).
		
		\item $\nu=0$ -- $P_i$ is a center (not frequent). It is associated with free oscillations (center bottom panel in FIG. \ref{fig-clas-p-crit}).
		
		\item $\nu<0$ -- $P_i$ is a stable spiral (future attractor). The system does damped oscillations until it settles down in the equilibrium state (right-hand bottom panel in FIG. \ref{fig-clas-p-crit}).
		
		\end{enumerate} 

\end{enumerate} 

The above classification of isolated critical points encompasses the kinds of equilibrium points most frequently encountered in cosmological applications. Any curve in the phase space which connects critical points (two or more, perhaps all of them) is called as ``heteroclinic orbit.'' A curve which joints a critical point with itself is called ``homoclinic trajectory.'' There are other useful concepts which are not explained here. The interested reader is submitted to the bibliography \cite{brauer, arrowsmith, hirsh, perko, arnold, ellis}.

\end{document}